\documentclass[prb,a4paper,10pt,twocolumn,showpacs,floatfix,superscriptaddress,amsmath,amsfonts,amssymb,preprintnumbers,longbibliography,citeautoscript]{revtex4-1}
\usepackage[T1]{fontenc}
\usepackage[utf8]{inputenc} 
\usepackage{dcolumn,graphicx,color,booktabs,microtype,afterpage}
\graphicspath{{./}{figure/}}
\usepackage[charter,greekuppercase=italicized]{mathdesign}\usepackage{sidecap}

\renewcommand{\figurename}{Fig.}
\renewcommand{\tablename}{Table}
\makeatletter\renewcommand{\fnum@figure}[1]{\figurename~\thefigure~(Color online)\ }\makeatother
\makeatletter\renewcommand{\fnum@table}[1]{\tablename~\thetable.}\makeatother

\newcount\hh \newcount\mm
\hh=\time \divide\hh by 60
\mm=\hh \multiply\mm by 60 \mm=-\mm
\advance\mm by \time
\def\now{\number\hh:\ifnum\mm<10{}0\fi\number\mm}

\begin{document}

\makeatletter\renewcommand{\ps@plain}{%
\def\@evenhead{\hfill\itshape\rightmark}%
\def\@oddhead{\itshape\leftmark\hfill}%
\renewcommand{\@evenfoot}{\hfill\small{--~\thepage~--}\hfill}%
\renewcommand{\@oddfoot}{\hfill\small{--~\thepage~--}\hfill}%
}\makeatother\pagestyle{plain}


\title{~\vspace{-1ex}\\
Pairing of weakly correlated electrons in the platinum-based\\%
centrosymmetric superconductor SrPt$_3$P }

\author{T.\,Shiroka}\email[Corresponding author: \vspace{8pt}]{tshiroka@phys.ethz.ch}
\affiliation{Laboratorium f\"ur Festk\"orperphysik, ETH H\"onggerberg, CH-8093 Z\"urich, Switzerland}
\affiliation{Paul Scherrer Institut, CH-5232 Villigen PSI, Switzerland}

\author{M.\,Pikulski}
\affiliation{Laboratorium f\"ur Festk\"orperphysik, ETH H\"onggerberg, CH-8093 Z\"urich, Switzerland}

\author{N.\,D.\,Zhigadlo}
\affiliation{Laboratorium f\"ur Festk\"orperphysik, ETH H\"onggerberg, CH-8093 Z\"urich, Switzerland}

\author{B.\,Batlogg}
\affiliation{Laboratorium f\"ur Festk\"orperphysik, ETH H\"onggerberg, CH-8093 Z\"urich, Switzerland}

\author{J.\,Mesot}
\affiliation{Laboratorium f\"ur Festk\"orperphysik, ETH H\"onggerberg, CH-8093 Z\"urich, Switzerland}
\affiliation{Paul Scherrer Institut, CH-5232 Villigen PSI, Switzerland}

\author{H.-R.\,Ott}
\affiliation{Laboratorium f\"ur Festk\"orperphysik, ETH H\"onggerberg, CH-8093 Z\"urich, Switzerland}
\affiliation{Paul Scherrer Institut, CH-5232 Villigen PSI, Switzerland}

\begin{abstract}
\noindent 
We report a study of the normal- and superconducting-state electronic properties of the centrosymmetric 
compound SrPt$_3$P via ${}^{31}$P nuclear-magnetic-resonance (NMR) and magnetometry investigations. 
Essential features such as a sharp drop of the Knight shift at $T < T_c$ and 
an exponential decrease of the NMR spin-lattice relaxation ratio $1/(T_1T)$ below $T_c$ 
are consistent with an $s$-wave electron pairing in SrPt$_3$P, although a direct 
confirmation in the form of a Hebel-Slichter-type peak is lacking. 
Normal-state NMR data at $T < 50$\,K indicate conventional features of the conduction electrons, 
typical of simple metals such as lithium or silver.
Our data are finally compared with available NMR results for the noncentrosymmetric 
superconductors LaPt$_3$Si and CePt$_3$Si, which adopt similar crystal structures.
\end{abstract}



\maketitle\enlargethispage{3pt}

\vspace{-5pt}\section{Introduction}\enlargethispage{8pt}
The continuing search for new superconductors has recently resulted in the identification of a new family of 
superconducting phosphide compounds with the chemical composition \textit{A}Pt$_3$P (\textit{A} = Sr, Ca, or La).\cite{Takayama12} 
These materials, whose superconductivity has been claimed to be driven by conventional electron-phonon 
interactions, adopt a distorted anti-perovskite structure, resembling the structure 
of several noncentrosymmetric superconductors, such as LaPt$_3$Si or 
CePt$_3$Si.\cite{Bauer2005}
Contrary to the latter, however, 
SrPt$_3$P exhibits an inversion center due to the staggered arrangement of Pt octahedra.
Non\-centrosymmetric superconductors, characterized by antisymmetric spin-orbit couplings, are currently 
the subject of intense research.\cite{Yogi04,Yogi06,Mukuda09,Bauer12} 
Therefore, a comparative study of their centrosymmetric counterparts is of particular interest.
Indeed, the Sr-based compound of the new family, with a critical temperature $T_c = 8.4$\,K, has 
already been investigated in some detail. 
The first study included measurements of the specific heat  $C(T)$ and Hall resistivity 
$\rho_\mathrm{H}(T)$.\cite{Takayama12} 
The specific-heat data were interpreted as indicating a strong electron-phonon coupling with a relatively large 
ratio $2 \Delta_0 / k_{\mathrm{B}}T_c \sim 5 $, and a fully-gapped excitation spectrum below $T_c$ with 
a zero-temperature gap value $\Delta_0 = 1.85$ meV. The nonlinear magnetic field dependence of 
Hall resistivity was attributed to the presence of multiple Fermi-surface pockets.
Less details are known on LaPt$_3$P, whose critical temperature $T_c \simeq 1.5$\,K\cite{Takayama12} 
is distinctly lower than that of the Sr compound, the difference being most likely due to the 
different valencies of the cations.

There exist several theoretical interpretations of the SrPt$_3$P data, which differ in their conclusions.
In Ref.~\onlinecite{Chen12}, 
the proximity to a dynamical charge-density wave instability is claimed to favor the occurrence of 
superconductivity in SrPt$_3$P.
In another approach, based on a Mig\-dal-Elia\-sh\-berg-ty\-pe analysis, it is argued that a conventional 
phonon-mediated superconductivity is observed.\cite{Subedi13} 
Density-functional-theory calculations\cite{Kang13}
suggest that the onset of superconductivity is due to a strong coupling between the $pd\pi$-hybridized bands with 
low-energy phonon modes confined in the $ab$ plane of the crystal lattice. The 2D character of these modes seems 
essential to preserve the antipolar arrangement of the distorted Pt octahedra in SrPt$_3$P, thus enhancing both the electron-phonon coupling 
constant $\lambda_{\mathrm{ep}}$ and, consequently, the critical temperature $T_c$. 

The only experimental investigation at a microscopic level cited in the literature focused on muon-spin rotation 
($\mu$SR) measurements.\cite{Khasanov14} The established temperature dependence of the penetration depth 
in the form of $\lambda^{-2}(T)$ again indicated a single gap with a value of  $\Delta_0 = 1.55$\,meV.  However, 
an upward curvature of the upper critical field $H_{c2}(T)$ just below $T_c$ was interpreted as reflecting two-band 
superconducivity with equal gaps but differing values of the coherence lengths.

In this work we report on the microscopic physical properties of SrPt$_3$P, 
investigated by means of nuclear magnetic resonance.\cite{Walstedt08,Curro09}
The measurements described below are based on $^{31}$P NMR, and 
probe both static (line widths and -shifts), as well as dynamic (spin-lattice 
relaxation) properties of the material. Overall the data indicate that 
SrPt$_3$P is a rather simple metal and that the pairing configuration 
is most likely of spherically-symmetric ($s$-wave) character.
A comparison with the non\-centro\-symmetric superconductors LaPt$_3$Si and CePt$_3$Si 
is intended to clarify the peculiarities of the latter. The lack of inversion symmetry of the 
crystal lattice may partly account for the much lower $T_c$ of LaPt$_3$Si, while the 
rare-earth cations seem responsible for the unusual electronic properties of 
CePt$_3$Si.\cite{Yogi04,Mukuda09}

\vspace{-5pt}
\section{Experimental details\label{sec:methods}}
\begin{figure}[t]
\includegraphics[width=0.85\columnwidth]{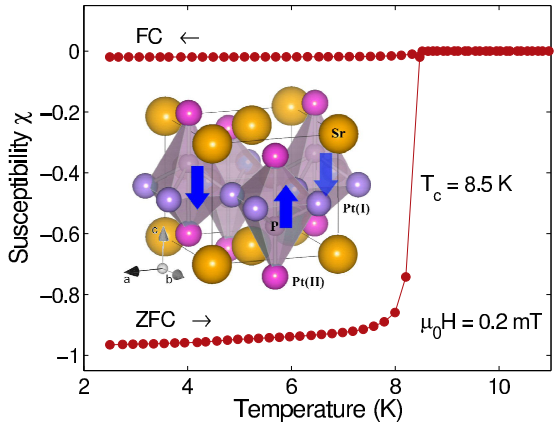} 
\caption{\label{fig:magnetization}Temperature dependence of the zero field-cooled (ZFC) and field-cooled (FC) dc magnetization measured at  $\mu_0H = 0.2$ mT. The sample shows a sizeable 
diamagnetic response with $T_c = 8.5$\,K.
Inset: structure of SrPt$_3$P showing the Pt$_6$ octahedra 
(adapted from Ref.~\onlinecite{Kang13}). The asymmetric positions of the apical Pt(II) ions 
give rise to alternating electric polarization vectors (blue arrows), hence preserving the 
centrosymmetric nature of  SrPt$_3$P.
}
\end{figure}

Polycrystalline samples of SrPt$_3$P were synthesized under conditions of high 
pressure and high temperature. High-purity (99.99\%) coarse powders of Sr, Pt, 
and P were mixed in the stoichiometric 1:3:1 ratio, then thoroughly ground, and 
finally enclosed in a boron-nitride container. To avoid a possible degradation 
of the material due to air exposure, all the preparatory steps were done in a 
glovebox under argon atmosphere. For the heat treatment, the BN crucible 
was placed into a pyrophyllite cube containing a graphite heater. An external 
pressure of 2 GPa was applied at room temperature and kept constant during 
the two-hour ramping up of the temperature to 1050$^{\circ}$C. The final 
temperature was kept constant for 40 hours and then lowered back to room 
temperature within approximately one hour. After releasing the pressure, the 
polycrystalline samples were extracted and kept in closed capsules filled with 
Ar gas during the corresponding experiments. 
Subsequently-recorded x-ray powder diffraction patterns agreed with those 
reported in Ref.~\onlinecite{Takayama12}. Measurements of the low-temperature 
magnetization revealed the onset of a diamagnetic response at $T_c = 8.5$\,K, 
again in good agreement with the originally reported value.\cite{Takayama12} 
The relevant magnetic susceptibility data $\chi(T)$ are shown in Fig.~\ref{fig:magnetization}  
where, at low temperatures, we observe a significant diamagnetic response, 
typical for the onset of superconductivity.\footnote{Since the magnetic response of a 
superconductor is influenced by many factors, such as the sample granularity, its compactness, 
flux pinning, the demagnetization factor, the remnant field, etc., the proximity 
to the ideal diamagnetic response is rather coincidental in our case.}
%
%

Although all the constituent elements of SrPt$_3$P feature NMR-active nuclei, the  ${}^{31}$P nucleus is clearly 
the best choice with respect to isotopic abundance (100\%) and gyromagnetic ratio ($\gamma/2\pi = 17.2356$ MHz/T). 
In addition, the $I = \frac{1}{2}$ nuclear spin of  ${}^{31}$P helps avoiding 
complications due to quadrupole effects.
In order to probe an adequate region of the superconducting part of the $H$-$T$ phase diagram above 1.6 K 
(i.e., accessible with a $^4$He flow cryostat), 
the external field for the NMR studies was set to  $\mu_0 H_{\mathrm{ext}} = 2$ T, corresponding 
to a critical temperature of approximately 5\,K (see, e.g., Fig.~1 in Ref.~\onlinecite{Khasanov14}). 
Analogous measurements probing the normal state in the same temperature regime were made in a 
field of $\mu_0 H_{\mathrm{ext}} = 7$ T, which exceeds the zero-temperature 
upper critical field $H_{c2}(0) \simeq 5.5$\,T,\cite{Khasanov14} 
and, for monitoring the superconducting transition, in a 0.6-T field  
across a restricted temperature range around the relevant $T_c$.

In order to achieve an acceptable signal-to-noise (S/N) ratio, 
the sample was powdered to a grain size of the order of 10\,$\mu$m and kept 	 
in the form of a loose powder, hence reducing the electrical contact between the grains. 
NMR measurements at the above-mentioned external magnetic fields were made 
at different temperatures between 1.6 and 300\,K by employing a conventional 
phase-coherent spectrometer. 
The $^{31}$P spectra were obtained by fast Fourier transformation (FFT) of the spin-echo signals 
generated by $\pi/2 - \pi$ rf pulses with a 200-$\mu$s delay between pulses (the measured 
transverse relaxation time is $T_2$ of ca.\ 500 $\mu$s). Since the employed ``hard'' rf pulses 
($t_{\pi/2} \sim 2.2$ $\mu$s) excite non-selectively the entire resonance line, no difference 
between fixed- and swept-frequency lineshapes was observed. The nuclear spin-lattice 
relaxation times $T_1$ were measured at the peak position of each NMR line by means of 
the inversion-recovery method, using spin-echo detection at variable delays.
The magnetic field was calibrated via ${}^{27}$Al NMR of elemental aluminum, 
whose gyromagnetic ratio and Knight shift are known to high precision. This data was 
subsequently used to calculate the ${}^{31}$P NMR shifts.

\vspace{-5pt}
\section{Experimental results and analysis}\label{sec:results}
\vspace{-5pt}
\subsection{NMR line shapes and shifts}
\begin{figure}[t]
\includegraphics[width=0.75\columnwidth]{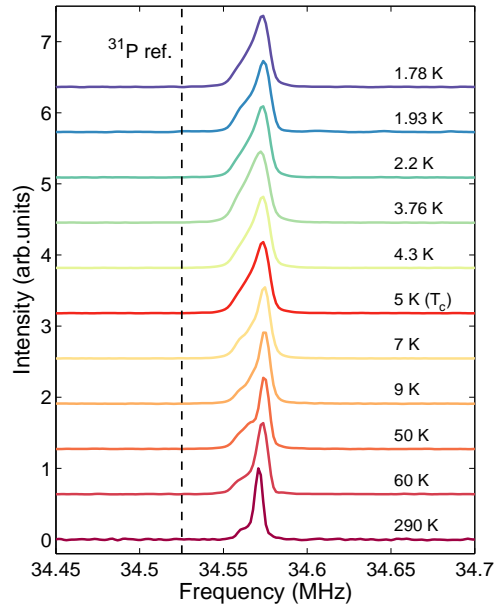} 
\caption{\label{fig:lines}Representative ${}^{31}$P NMR line shapes in SrPt$_3$P 
at $\mu_0H = 2.0$ T and temperatures in the range 1.75 to 290 K. The enhanced 
line widths and asymmetry below 5\,K reflect the onset of the superconducting phase.}
\end{figure}

In Fig.~\ref{fig:lines} we show the evolution of the  ${}^{31}$P lines with varying 
temperature at $\mu_0H = 2.0$ T, both above and below $T_c$. Similar results 
with respect to the line widths and -shifts were obtained for $\mu_0H = 7.0$\,T 
(see Fig.~\ref{fig:shifts_2_vs_7T}).

\begin{figure}[thb]
\includegraphics[width=0.9\columnwidth]{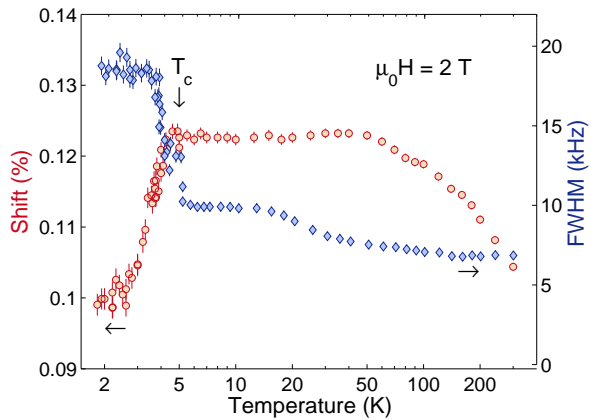} 
\caption{\label{fig:shifts}${}^{31}$P NMR shift (left scale) and linewidth (right scale) 
vs.\ temperature, as measured at an applied field of 2\,T. 
The drop in shift and the increase in width below $T_c$ are followed by a constant behavior 
at intermediate temperatures and by a final decrease at higher $T$ (see text for details).
The increase of data scattering below $T_c$ reflects the decreased S/N ratio 
in the superconducting phase.}
\end{figure}

We focus first on the normal phase. 
As can be seen in Fig.~\ref{fig:shifts}, the line width, of the order of 7\,kHz at 
room temperature, increases gradually by approximately 50\% down to ca.\ 12\,K 
and subsequently it stays constant between 12\,K and $T_c$. 
As for the line shifts in the normal phase, a moderate increase upon lowering the 
temperature is observed.
This increase is mostly linear down to approximately 50\,K, below which the line 
shift remains constant down to $T_c$ (see inset of Fig.~\ref{fig:shifts_2_vs_7T}).
The constancy of the shift above the superconducting transition is a feature often observed in simple metals 
with an only weakly-varying electronic susceptibility $\chi_{\mathrm{P}}(T)$. The relative shift, of the 
order of 0.12\%, is close to that observed in simple metals such as Ag and Na and is the same for 
the 2- and 7-T external fields, respectively (see Fig.~\ref{fig:shifts_2_vs_7T}). The line widths measured in a 
field of 7\,T, instead, are approximately twice as large as those recorded at $\mu_0H = 2$\,T, 
which, in turn, are three times larger than those at 0.6\,T. Thus the linewidth 
scales almost linearly with the applied field.

The most interesting results are obtained below $T_c$, where the inhomogeneous 
field distribution in the mixed state of a type-II superconductor significantly broadens 
the NMR line. Fig.~\ref{fig:shifts} shows the steep increase of the line width, from 10 
to ca.\ 18 kHz, between $T_c$ and 3.5\,K (and from 3.2 to 8\,kHz at 
0.6\,T - data not shown). 
Upon a further reduction of $T$, the width remains constant down to the lowest 
reached temperatures. At the same time the $^{31}$P Knight shift exhibits a 
sudden decrease below $T_c$, a strong indication of spin-singlet superconductivity, 
since the pairing of electrons with opposite spins occurring in type $s$- (or $d$-wave) 
superconductors implies a significant decrease of the local spin susceptibility, as detected by 
the probe nuclei. Of the two possibilities, the experimental data rule out the occurrence 
of a $d$-wave SC pairing, since in that case, the drop of the Knight shift would be 
less abrupt and the line broadening would reach its asymptotic $T = 0$\,K value 
only gradually.\cite{Walstedt08} 
From these static NMR results, an $s$-type pairing seems to be the most plausible 
configuration adopted by electrons in the superconducting state of SrPt$_3$P.

\vspace{-5pt}
\subsection{NMR relaxation rates and nature of superconductivity}
The conclusion about the $s$-wave pairing configuration is also supported by the 
$^{31}$P spin-lattice $1/T_1$ relaxation-rate data displayed in Fig.~\ref{fig:relaxations}.
The $1/(T_1T)$ vs.\ $T$ plot exhibits what is considered to be a typical 
signature of conventional BCS superconductivity, an exponential decrease with decreasing 
$T$ well within the superconducting phase, but no clear indication of a Hebel-Slichter-type 
coherence peak, which is masked by a distinct increase of $1/(T_1T)$  in a narrow regime 
just above $T_c$.
%

%
\begin{figure}[b]
\includegraphics[width=0.8\columnwidth]{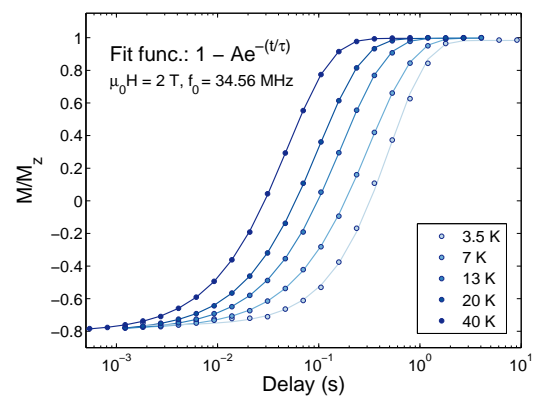} 
\caption{\label{fig:T1_curves_2T}${}^{31}$P NMR spin-lattice relaxation data of 
SrPt$_3$P at selected temperatures measured at 2\,T. All fit curves exhibit a simple 
exponential ($\beta = 1$) recovery of the magnetization (see text).}
\end{figure}
The spin-lattice relaxation times $T_1$ were evaluated from magnetization-recovery 
curves such as those shown in Fig.~\ref{fig:T1_curves_2T}. The nuclear magnetization data 
$M_z(t)$ were fitted by using the  single-exponential equation valid for spin-1/2 nuclei:
\begin{equation}
\label{eq:T1-IR}
M_z(t) = M_z^0 \left[1-f\exp(-t/T_1)^{\beta}\right].
\end{equation}
Here $M_z^0$ represents the saturation value of magnetization at thermal equilibrium,
$f$ is the inversion factor (exactly 2 for a complete inversion), and $\beta$ is a 
stretching exponent which accounts for possible distributions 
of relaxation rates.\cite{Shiroka2011}
Except at the lowest temperatures, the $\beta$ values were always found to be 
close to 1, with no difference in fit quality whether $\beta$ was fixed or set free. 
Considering the purely magnetic relaxation of $I=1/2$ nuclear probes, this implies 
the same local environment for all the $^{31}$P nuclei, i.e., disorder of either 
structural or magnetic nature is insignificant in our case.

\begin{figure}[t]
\includegraphics[width=0.9\columnwidth]{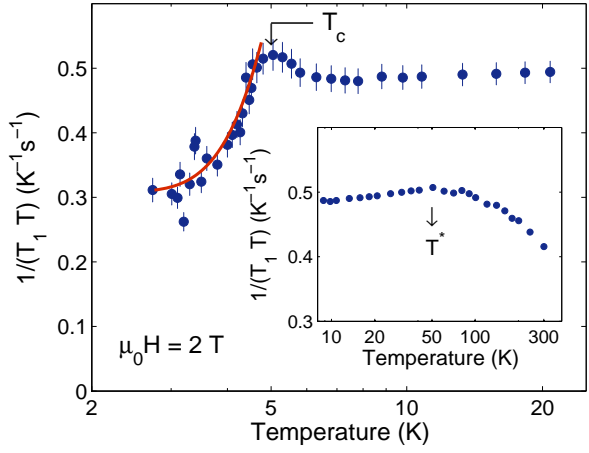} 
\caption{\label{fig:relaxations}Temperature dependence of $(T_1T)^{-1}$ measured 
in a magnetic field of 2\,T. A similar cusp-like feature at $T_c$ is seen also at 0.6\,T.
The decrease below the maximum occurring at $T_c$ 
is compatible with 
a fully-gapped superconductor (see text). Inset:  $(T_1T)^{-1}$ data 
over the full temperature range show a broad maximum at $T^*=45$\,K 
(see also Fig.~\ref{fig:suscept}).} 
\end{figure}

Up to $T^* = 45$\,K, the  normal-state spin-lattice relaxation rate varies approximately linearly with 
temperature (see inset in Fig.~\ref{fig:relaxations}), as is usually observed 
in conventional simple metals. 
The progressive deviation from the linear-in-$T$ behavior above $T^*$ reflects a 
decrease in the dynamical spin susceptibility, as discussed in Sec.~\ref{ssec:korringa} 
and \ref{ssec:susceptibility}.

Focusing back on the superconducting state, just below $T_c$ we do not 
observe a distinct Hebel-Slichter (coherence) peak, but only a cusp-like feature peaking at $T_c$. 
Analogous data taken in a 0.6-T field showed essentially the same type of feature in the vicinity 
of $T_c$. 
%
We recall that 
the Hebel-Slichter anomaly is due to pair coherence and to a quasiparticle density-of-states anomaly 
at the gap edge and is more pronounced at lower fields, where the pair-breaking effect of the field 
is less important.\cite{Masuda69} %
Both these effects can be weakened by different causes, including interaction 
anisotropies and the broadening of the quasiparticle states due to strong-coupling effects.
Indeed, while \textit{weakly}-coupled conventional $s$-wave superconductors, 
typically exhibit a Hebel-Slichter peak, this is not the case in the \textit{strong}-coupling limit, 
where the enhanced electron-phonon coupling implies reduced quasiparticle lifetimes.\cite{Ohsugi92}
%

Since the lack of a coherence peak cannot rule out 
the possibility of an $s$-type electron pairing,\cite{Parker08} further analysis is required.
%
%
The temperature dependence of $T_1$ at $T < T_c$ can provide information about the symmetry 
of the superconducting gap. In case of an anisotropic gap with nodes, a power-law dependence, $1/T_1 \sim T^{n}$, 
is expected since very low-energy electronic excitations around gap nodes can still contribute to relaxation.
On the other hand, for standard fully-gapped superconductors with no or a weak gap anisotropy, 
the relaxation rate depends 
exponentially on temperature, $1/T_1 \sim \exp(-\Delta_0/T)$, with $\Delta_0$ the 
superconducting gap at $T = 0$.
In either case, non-negligible $1/(T_1T)$ residual values may be observed at the lowest 
temperatures.\footnote{Since samples of different quality show similar 
residual values, the observed offset in $1/(T_1T)$ is likely of intrinsic origin.}
Most likely these are due to the presence of vortex-core or of thermally-excited vortex-motion 
relaxation mechanisms,\cite{MacLaughlin76,Corti96} which can be accounted for by an 
additional constant term. 
While both models provide reasonable fits to the data, the power-law expression 
shows a somehow poorer fit quality and, most importantly, the resulting exponent varies significantly 
(from 3.6 for the 2-T dataset to 2.7 for the data collected at 0.6\,T). 
On the other hand, as shown in  Fig.~\ref{fig:relaxations}, an exponential curve with 
$\Delta_0 = 1.47(4)$\,meV fits the SrPt$_3$P relaxation data satisfactorily well. Also the data 
taken at 0.6\,T can be consistently fitted with a similar gap value, $\Delta_0 = 1.51(4)$\,meV, 
thus providing a good indication 
that the superconducting gap of SrPt$_3$P is nodeless.
The value of the gap parameter, 
although slightly modified by the presence of the constant term, 
is in fair agreement with the 1.55\,meV value reported in 
Ref.~\onlinecite{Khasanov14}, but distinctly lower than $\Delta_0 = 1.85$\,meV reported in 
Ref.~\onlinecite{Takayama12}.
In any case, since the gap value from our analysis is clearly higher than 1.27 meV 
(the gap expected from the BCS formula $2 \Delta_0 = 3.52 k_{\mathrm{B}}T_c$), 
it is compatible with a strong-coupling scenario.

Besides the vortex-core relaxation mechanism mentioned above, 
the almost constant value 
of $1/(T_1T)$ well below $T_c$ 
might also reflect a certain degree of superconducting-gap anisotropy, but we 
consider this as a less likely possibility.

\vspace{-5pt}
\subsection{Korringa relation and the degree of electronic correlation\label{ssec:korringa}}
Some insight into the degree of electronic correlations in the normal state of a 
material can be gained by considering the so-called Korringa relation\cite{Korringa50}
\begin{equation}
\label{eq:Korringa_r}
T_1TK_s^2 = S_0, \quad \text{with} \quad 
S_0 = \frac{\gamma_e^2}{\gamma_n^2} \frac{\hbar}{4 \pi k_{\mathrm{B}}},
\end{equation}
Here $\gamma_e$ and $\gamma_n$ are the electronic and the nuclear gyromagnetic ratios, respectively, 
while $K_s$ is the line shift due to the polarized conduction electrons. Eq.~(\ref{eq:Korringa_r}) 
reflects the fact that both the readily accessible experimental parameters, 
the Knight shift $K_s$ and the spin-lattice relaxation time $T_1$, depend on the same 
electron-nucleus hyperfine interaction (assumed to be mostly of Fermi-contact character). 
However, while for a simple (Fermi-gas) metal the parameter $S_0$ is a constant (depending 
only on $\gamma_n$ of the probe-nucleus), for real materials the generalized 
relation\cite{Narath68} $T_1TK_s^2 = \alpha S_0$, with $\alpha$ a measure of 
the strength of the (Fermi-liquid) quasiparticle interactions,\cite{Pines54} is more appropriate.
Since $1/(T_1T)$ probes the dynamical susceptibility averaged over the Brillouin zone, either 
FM or AFM spin correlations will enhance it, but only FM correlations can significantly enhance 
also the shift. 
Thus, $\alpha > 1$ ($< 1$) indicates the presence of ferro- (antiferro-)magnetic electronic correlations.

From the almost constant value of $S_0/(T_1TK_s^2)$ vs.\ $T$
(see inset in Fig.~\ref{fig:relaxations}) and 
considering the negligible orbital shift contribution (see below), we find 
$1/\alpha \simeq 0.55$ and, hence, $\alpha = 1.8$ in our case. 
The resulting value, slightly larger than unity, seems to suggest 
the presence of ferromagnetic correlations in the normal state. However, the ratio 
$1/\alpha \simeq 0.55$ is similar to 0.58, 053, and 0.50 observed in typical 
metals such as lithium, cesium, and silver, respectively.\cite{Walstedt08} In these nearly-free-electron 
metals the Moriya theory of exchange enhancement, whereby the exchange fluctuations 
also enhance the rate of the $T_1$ process,\cite{Moriya63,Walstedt08} explains the 
experimental data satisfactorily. Consequently, we conclude that also in SrPt$_3$P the conduction 
electrons experience rather weak correlation effects, in good agreement with 
the results reported in Ref.~\onlinecite{Takayama12}. 

\begin{figure}[t]
\includegraphics[width=0.9\columnwidth]{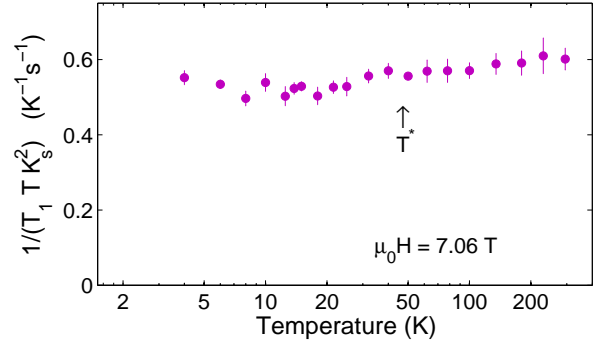} 
\caption{\label{fig:korringa}Korringa relation [reported in 
$S_0$ units, see Eq.~(\ref{eq:Korringa_r})] shows only a weak temperature dependence, 
with no distinct features at $T^*=45$\,K (see also Fig.~\ref{fig:suscept}).}
\end{figure}

\vspace{-5pt}
\subsection{Magnetic susceptibility and NMR\label{ssec:susceptibility}}
\begin{figure}[t]
\includegraphics[width=0.9\columnwidth]{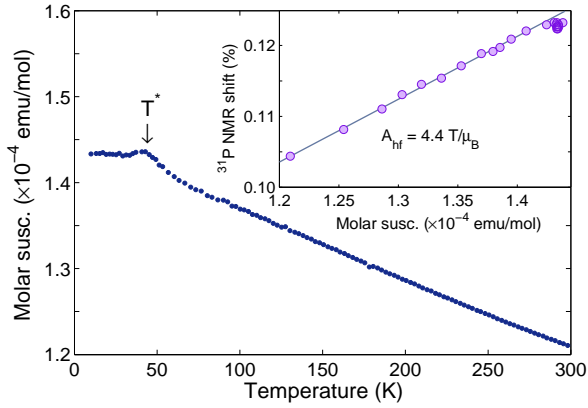} 
\caption{\label{fig:suscept}Temperature dependence of the SrPt$_3$P molar magnetic 
susceptibility $\chi_{\mathrm{mol}}$ in an applied field of 2\,T. Data were corrected by the 
addition of $1.18\times10^{-4}$\,emu/mol to account for the core diamagnetism 
(see text). Inset: ${}^{31}$P Clogston-Jaccarino ($K$-$\chi$) plot at 
$\mu_0 H = 2$\,T. The line is a fit to the $T > 50$\,K data.}
\end{figure}

The magnetization $M(T)$ was measured in an applied magnetic field of 2 T in specimens from the 
same batch as that used for NMR experiments. The resulting values of the susceptibility, 
$\chi = M/H$, for temperatures above 10\,K are shown in the main panel of Fig.~\ref{fig:suscept}.
The measured net susceptibility seems unusually small for a metal, because the contribution 
from the conduction electrons is almost exactly canceled by the large core diamagnetism mainly 
from Pt ($1.18\times10^{-4}$\,emu/mol per formula unit, according to standard tables). This 
compensation effect makes it possible to measure even subtle variations of the electronic 
contribution with $T$.

\begin{figure}[t]
\includegraphics[width=0.8\columnwidth]{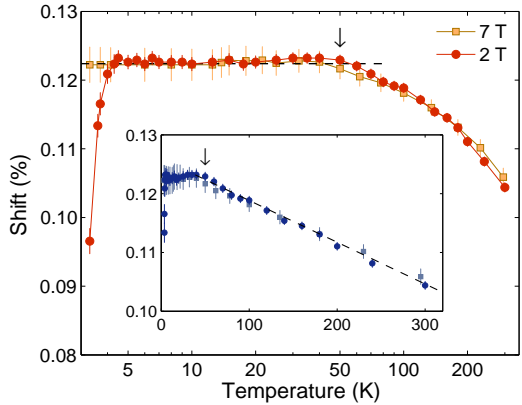} 
\caption{\label{fig:shifts_2_vs_7T}Comparison of ${}^{31}$P NMR line shifts vs.\ 
temperature at 2 (circles) and 7\,T (squares) on logarithmic (main panel) and on 
linear scale (inset). The nature of the shift changes from constant to linearly 
decreasing at 45 K (arrows). The dashed lines highlight both regimes and serve 
as guides to the eye.}
\end{figure}

In order to relate $\chi(T)$ with the NMR data, the relative ${}^{31}$P 
NMR shifts in the two chosen magnetic fields (2 and 7\,T, respectively)
are shown in Fig.~\ref{fig:shifts_2_vs_7T}. 
As expected for simple metals, the shifts are independent of the applied field and hence 
coincide perfectly. Below $T_c$ the high-field (7-T) data reflect the normal phase: both 
the line shift and width remain practically constant upon decreasing temperature. 

From the $\chi(T)$ data (shown in Fig.~\ref{fig:suscept}) and the NMR shifts (see inset 
of Fig.~\ref{fig:shifts_2_vs_7T}), one recognizes immediately very similar features.
In either case, a practically linear increase upon lowering the temperature,  
from 300\,K down to ca.\ $T^* = 45$\,K, is followed by a $T$-independent part.
This clear break in the temperature dependence of both $K_s(T)$ and $\chi(T)$ 
coincides with a maximum of the Hall constant $R_{\mathrm{H}}$ reported in Ref.~\onlinecite{Takayama12}.
The reason for this might be that the trend to $R_{\mathrm{H}} = -1/ne$ starts to dominate  
the $R_{\mathrm{H}} > 0$ contribution. 

Such a comparison of micro- (NMR) vs.\ macroscopic (magnetometry) results provides 
further information about the normal-state electronic properties of SrPt$_3$P. In our case, given 
the relative insensitivity to the applied field (see Fig.~\ref{fig:shifts_2_vs_7T}), 
it makes sense to compare the NMR shifts with the susceptibility data, as shown in 
the Clogston-Jaccarino\cite{Clogston64} (or $K$-$\chi$) plot in the inset of 
Fig.~\ref{fig:suscept}. Here both data sets refer to the same (2-T) field, with 
the temperature as an implicit parameter.
From the slope $\mathrm{d}K/\mathrm{d}\chi$ of the $K$-$\chi$ plot for $T > 50$\,K, 
and by assuming a standard electronic $g$-factor of 2.0, the relation 
$K = g A_{\mathrm{hf}} \chi + K_{\mathrm{orb}}$ gives the hyperfine coupling 
constant, $A_{\mathrm{hf}} = 4.4$\,T/$\mu_{\mathrm{B}}$, 
while the zero intercept with the $K$-axis gives the orbital shift, $K_{\mathrm{orb}} = -0.003$\%. 
The latter is perfectly compatible with known orbital-shift values for 
a light nucleus such as ${}^{31}$P. 
Clearly, the orbital shift is negligible in comparison with the dominant magnetic shift.
We recall that hyperfine coupling constants are known to be strongly material-dependent, so it 
is not unusual that our value is an order of magnitude higher than, e.g., 
$A_{\mathrm{hf}} \sim 0.2$\,T/$\mu_{\mathrm{B}}$ found in another 
phosphorus-based compound, Pb$_2$(VO)(PO$_4$)$_2$.\cite{Nath09}

\vspace{-5pt}
\section{Discussion}\label{sec:discussion}
Below we discuss in more detail some aspects of the results presented in the previous sections.

\emph{Line shapes.} 
For all the applied magnetic fields the normal-state ${}^{31}$P lines show a slight 
asymmetry, usually attributed to the crystallite surfaces or to minor impurities.
However, the persistence of the line asymmetry even in samples of different quality, 
most likely indicates a (uniaxial) anisotropy of the hyperfine coupling tensor 
(whose study is beyond the scope of this work), which can be
observed also in powdered samples.\cite{Kawamoto95}
The uniaxial nature of the HF tensor is to be expected, if the symmetry 
of the distorted Pt octahedra probed by the phosphorus nuclei is considered 
(see structural unit in the inset of Fig.~\ref{fig:magnetization}).

The line shapes, in either the normal or the superconducting phase, indicate 
also the structural and chemical homogeneity 
of the sample. 
An essentially single line with a relatively narrow width of only 
$\sim 7$\,kHz (see Fig.~\ref{fig:lines}) is typical of defect-free crystal structures, 
especially when considering the random polycrystalline nature of the sample.
Similarly, the line asymmetry observed below $T_c$, again reflects the high 
quality of the sample, since in a powdered superconductor with a significant 
degree of disorder the lines would appear symmetric.

\emph{Residual shift.} 
A key prediction of the BCS theory for superconductors with singlet $s$-wave pairing 
is the vanishing of the Knight shift for $T \ll T_c$, reflecting the vanishing 
susceptibility due to the pairing of quasiparticles with antiparallel spins. Yet, even for elemental superconductors 
(as, e.g., Sn or Hg), a residual shift is observed even close to $T = 0$.\cite{MacLaughlin76}
The presence of a significant orbital Knight shift and/or of spin-orbit scattering (known as Ferrell-Anderson effect) 
have been identified 
as responsible for its occurrence.
The SrPt$_3$P case, however, is a bit more complex, since here $^{31}$P is a low-$Z$ 
nucleus 
which interacts with superconducting carriers arising mostly from in-plane 
$pd\pi$-hybridized states between Pt and P ions.\cite{Kang13}
Since orbital shifts are negligible for the light probe-nuclei,  
the presence of a low-temperature nonzero Knight shift in SrPt$_3$P 
is, therefore, most likely due to spin-orbit scattering effects.

\emph{Core-polarization effects and $T^*$ anomaly.} 
Another contribution to the nonvanishing shift at $T = 0$ can arise from 
core-polarization effects,\cite{Freeman65,MacLaughlin76} whereby the exchange 
interaction of localized core electrons with field-polarized conduction band electrons 
(partially of $d$ character in our case), implies an additional nonzero contact interaction 
in the $^{31}$P nuclei. 
The observation of large negative  $^{195}$Pt-NMR Knight shifts (data not shown) 
seems to confirm the presence of core-polarization effects in SrPt$_3$P.\cite{Clogston64}
While this contribution to the Knight shift would disappear entirely if purely $d$-band electrons 
were coupled into Cooper pairs, a residual shift would still arise considering the hybridized 
(i.e., not exclusively $d$) character of the conduction band quasiparticles.

The consequences of core-polarization effects are more important, however, in the 
normal state. The electronic contribution to the magnetic susceptibility of a $d$-type 
band is generally temperature dependent, thus implying a temperature-dependent 
Knight shift, as actually observed (for an ideal metal 
one would have expected a constant shift). While core-polarization effects might explain 
the similar $T$-dependence of $K$ and$\chi_s$, the abrupt change in slope observed at $T^*$, 
as well as the value of $T^*$ itself, remain to be investigated. 

\emph{Weak electron correlation.} 
Our NMR data of SrPt$_3$P, indicate a fairly simple metallic behavior in the normal state, 
followed by standard BCS features in the superconducting state. These results are 
in remarkable agreement with the transport and magnetometry measurements reported in 
Ref.~\onlinecite{Takayama12} in two important aspects. 
Firstly, both NMR (via the Korringa product) and the macroscopic measurements (via the 
Sommerfeld-Wilson ratio\footnote{The Sommerfeld-Wilson ratio, the dimensionless ratio 
of the zero-temperature magnetic susceptibility and the linearly varying term of the 
temperature dependence of specific heat, can provide insight into the degree of 
electronic correlation.}) 
demonstrate the \textit{absence of significant electron correlations}.
Secondly, there is a distinct feature at $T^* = 45$\,K in the $^{31}$P Knight 
shift and in the $1/(T_1T)$ relaxation behavior, respectively, both of which 
change from a constant-in-$T$ to a linear-in-$T$ decrease as the temperature 
changes across $T^*$. 
This temperature coincides with the above-mentioned maximum in 
$R_{\mathrm{H}}(T)$, the latter reflecting anisotropies of the electron mean 
free path across different parts of the Fermi surface. 
The latter fact is consistent with conclusions drawn in Ref.~\onlinecite{Khasanov14}.
The coincidence of $T^*$ values, derived from results of very different types of 
experiments, is unclear at present.
 
As for the Korringa relation, we recall that the absolute value of $\alpha$ might also 
be influenced by extraneous factors, such as the presence of strong disorder,\cite{Shastry94} 
or effects related to the hyperfine form factor (i.e., $q$-space filtering). Nevertheless, as 
mentioned above, these can be excluded in our case and the observation of a constant 
Korringa ratio vs.\ temperature represents a valid proof of Fermi-liquid behavior in SrPt$_3$P, 
where $1/\alpha \simeq 0.55$ is constant over a fairly wide temperature range.

\emph{Centro- vs.\ noncentro-symmetric superconductors.} 
We conclude with a comparison of SrPt$_3$P with two noncentrosymmetric superconductors,
the non-magnetic LaPt$_3$Si ($T_c = 0.6$\,K) and the magnetic CePt$_3$Si ($T_c = 0.75$\,K), 
both of which were studied via NMR, as well.\cite{Yogi04,Yogi06,Mukuda09}
The latter not only share the same structure, but show also similar antisymmetric spin-orbit couplings. 
Hence, a comparison of the three is relevant to understand whether it is the lack 
of inversion symmetry, or the presence of rare-earth ions to account for the unusual 
properties of noncentrosymmetric superconductors.

While SrPt$_3$P and LaPt$_3$Si both follow a perfect Korringa law down to 
$T_c$, in CePt$_3$Si the $1/(T_1T)$ ratio is enhanced upon cooling due to 
the development of 4$f$-derived magnetic fluctuations.
Most importantly, unlike in SrPt$_3$P and LaPt$_3$Si, where the superconducting 
transition is of conventional type and not dominated by correlation effects, in 
CePt$_3$Si, the superconductivity emerges from a unique heavy-electron state 
which coexists with antiferromagnetic order. 

From these facts, we conclude that the key differences between the considered 
centro- and non-centrosymmetric superconductors, are most likely related to 
the presence of Ce$^{3+}$ ions and the ensuing heavy-fermion phenomena, 
while the parity mixing alone (common to both LaPt$_3$Si and CePt$_3$Si) 
cannot explain the peculiar behavior of the latter.
Despite the many similarities between SrPt$_3$P and LaPt$_3$Si, their belonging 
to different classes of crystal symmetry may possibly explain their distinctly 
different critical temperatures.
Incidentally, from the available data in literature, also the centrosymmetric LaPt$_3$P 
compound seems most likely to be a conventional BCS type-I superconductor.

\vspace{-5pt}
\section{Conclusions\label{sec:conclusions}}
We presented the results of ${}^{31}$P NMR measurements in the recently discovered 
SrPt$_3$P superconducting compound. In the normal state, the nuclear spin-relaxation 
rate obeys a Korringa-type relation $(T_1T)^{-1}_n = 0.522$ (sK)$^{-1}$ and 
the line shift is independent of temperature. In the superconducting state, 
the lack of a clear coherence peak in $T_1^{-1}$ just below $T_c$, 
most likely due to a strong electron-phonon coupling, is followed by an exponential, 
thermally-activated decrease upon further cooling. A similar behavior is shown also 
by the line shift.
The overall results of our NMR investigation make us conclude that SrPt$_3$P, which 
behaves as a standard metal in the normal state, is characterized by singlet pairing 
of the electrons, and a gap function with conventional $s$-wave-type symmetry.

Finally, a comparison with the analogous non-centrosymmetric superconductors 
LaPt$_3$Si and CePt$_3$Si, strongly suggests that the peculiarities of the latter 
are closely related to the presence of rare-earth cations. The lack of inversion 
symmetry may cause a much lower $T_c$, however.

\vspace{-5pt}
\section*{Acknowledgments}
\vspace{-5pt}

The authors thank R.\ Khasanov (Paul Scherrer Institut) for sharing his 
data prior to publication and for useful discussions. 
This work was financially supported in part by the Schweizerische Nationalfonds 
zur F\"{o}rderung der Wissenschaftlichen Forschung (SNF) and the NCCR research 
pool MaNEP of SNF.

\end{document}